\documentstyle[12pt]{article} 
\begin{document}
\begin{titlepage}
\title{A remark on Ma's "Cosmic gravitational background radiation"}
\author{Lajos Di\'osi\thanks{E-mail: diosi@rmki.kfki.hu}\\
KFKI Research Institute for Particle and Nuclear Physics\\
H-1525 Budapest 114, POB 49, Hungary\\\
{\it e-print archives ref.: quant-ph/9805077\hfill}}
\date{}
\maketitle
\begin{abstract}
The calculations of Guang-Wen Ma's recent Letter [Phys.Lett. A239, (1998)
209] contain an easily detectable error which makes his conclusions
irrelevant.
\end{abstract}
\end{titlepage}
Ma \cite{Ma98} has recently claimed that, starting with a reasonable
gravitational background radiation, he can deduce the spectrum of
space-time fluctuations of a certain model \cite{Kar66}, and he claims
that this spectrum is cosmologically sensible. However, the spectrum he
has deduced differs from the true spectrum of this model which I and
Luk\'acs \cite{DioLuk93} criticized as being cosmologically untenable.

Unfortunately, there is a crucial error in Ma's deduction, which can
easily be detected.

Ma's calculations start with a certain  \cite{fnote} plain wave
expansion of the gravitational radiation, see eqs.(10-21).
The amplitudes of the two independent polarizations of the plane wave
are denoted by $\delta$ and $C$, respectively.
The gravitational energy $W$, i.e. the volume integral of the r.h.s.
of eq.(19) is then expressed in terms of the author's amplitudes
$\delta$ and $C$.
I note here that the energy of plane waves should always be positive.
However, the Letter's energy expression (21) is indefinite: the terms
proportional to $C^2$ have been left out.

I have calculated the term missing from eq.(21):
$$
W_{(\beta)}(missing)=\frac{c^4 V}{32\pi G}\sum_k k^2 C_{(\beta)}^2.
$$
This term is absolutely relevant for what Ma is going to do later in
his Letter. The author's claim is based on the extrem smallness of a
certain function $D^2$ which governs the intensity of fluctuations.
It's bounded by $D^2_{up}\sim 10^{-12}$ as the Letter states. Now,
the missing term above, when added to Ma's incomplete expression (21),
amounts to the change $D^2\rightarrow 1+D^2$, as may be seen from
eq.(24). The smallness of $D^2$ disappears in the correct plane wave
expansion of the energy. This, in itself, invalidates the Letter's
claim.

\bigskip
I'm grateful to Philip Pearle for his important suggestions.
This work was supported by the Hungarian Scientific Research
Fund under Grant No. OTKA T016047.


\begin{thebibliography}{99}
\bibitem{fnote} Ma chooses a gauge which violates the spatial rotation
invariance, see Eq.(17b). This leads to the awkward forms
$\sum\small{\frac{1}{k}}\Pi k$ instead of 3-covariant expressions.
\end{thebibliography}
\end{document}